\documentclass[a4paper,UKenglish,cleveref, autoref, thm-restate]{oasics-v2021}

\title{Using Spoofax to Support Online Code Navigation} 

\author{Peter D. Mosses}{Delft University of Technology, The Netherlands \and Swansea University, Wales, United Kingdom \and \url{https://pdmosses.github.io/} }{p.d.mosses@tudelft.nl}{https://orcid.org/0000-0002-5826-7520}{}

\authorrunning{P.\,D. Mosses} 

\Copyright{Peter D. Mosses} 



\category{Development} 

\relatedversion{} 

\funding{The work reported here is part of the PLanCompS project, which was funded by an EPSRC grant to Swansea University from 2011 to 2015 (EP/I032495/1) in collaboration with colleagues at Royal Holloway University of London and City University in London.}

\acknowledgements{
Eelco Visser (1966–2022) was a visionary leader. For more than two decades, I~have benefited greatly from reading his persuasive publications, listening to his animated presentations, and discussing topics of common interest with him -- especially modular language specification. Since 2012, I~have also been an enthusiastic user of the Spoofax language workbench, developed by Eelco and his Programming Languages group at TU Delft. This paper reports on some work that I~carried out as a visitor at TU Delft since 2016. I am grateful to the reviewers of a previous version for helpful comments and suggestions.
}

\nolinenumbers 

\EventEditors{Ralf L\"{a}mmel, Peter D. Mosses, and Friedrich Steimann}
\EventNoEds{3}
\EventLongTitle{Eelco Visser Commemorative Symposium (EVCS 2023)}
\EventShortTitle{EVCS 2023}
\EventAcronym{EVCS}
\EventYear{2023}
\EventDate{April 5, 2023}
\EventLocation{Delft, The Netherlands}
\EventLogo{}
\SeriesVolume{109}
\ArticleNo{21}

\begin{document}

\maketitle

\begin{abstract}
Spoofax is a language workbench. A Spoofax language specification generally includes name resolution: the analysis of bindings between definitions and references. When browsing code in the specified language using Spoofax, the bindings appear as hyperlinks, supporting precise name-based code navigation. However, Spoofax cannot be used for browsing code in online repositories.

This paper is about a toolchain that uses Spoofax to generate hyperlinked twins of code repositories. These generated artefacts support the same precise code navigation as Spoofax, and can be browsed online. The technique has been prototyped on the CBS (Component-Based Semantics) specification language developed by the PLanCompS project, but could be used on any language after specifying its name resolution in Spoofax.
\end{abstract}


\section{Introduction}
\label{sec:Introduction}

Programming languages (including domain-specific languages (DSLs)) usually allow definitions of named entities, and references to entities via their names.

\subsection{Code Navigation}

In a GitHub Blog post \cite{Creager2021ISG}, Douglas Creager explains the concept of \emph{code navigation} based on names:
\begin{quote}
Code navigation is a family of features that let you explore the relationships in your code and its dependencies at a deep level. The most basic code navigation features are “jump to definition” and “find all references.” Both build on the fact that \emph{names} are pervasive in the code that we write.
\end{quote}
Although “jump to definition” sounds simple enough, the difficulty of its specification and implementation depends highly on the code language. The determination of the bindings between definitions and references is called name resolution; different languages often have quite different rules for it.

Name resolution can significantly complicate searching for the definition of a particular name when using an ordinary browser or editor. For example, a language may allow references to a name from other files than the file containing the definition; it may also allow the same name to be defined more than once, possibly with shadowing. For such languages, simple project-wide searches for the definition of a specific name, or for all references to that name, may be imprecise, and return unhelpful false positives.

Eelco Visser led the development of three declarative meta-languages for specifying name resolution: NaBL \cite{Konat2012DNB}, NaBL2 \cite{Neron2015TNR,Antwerpen2016CLS}, and Statix \cite{Pelsmaeker2019TLS,Rouvoet2020KWA,Antwerpen2018SAT,Antwerpen2021SSG}. The Spoofax language workbench \cite{Kats2010SLW,Konat2016BDM,Wachsmuth2014LDS} implements name resolution for any language whose rules are specified in one of these meta-languages. When browsing a file in the specified language, Spoofax equips each reference to a defined name with a hyperlink; clicking on the hyperlink moves the cursor directly to the referenced definition.
If the definition is in a different file from the reference, Spoofax automatically opens an editing window on that file. Similarly, Spoofax equips each definition with the list of hyperlinks to all the current references to it. These hyperlinks are precise, and provide reliable support for code navigation.

The languages whose name resolution rules have been specified and implemented in Spoofax include:
\begin{itemize}
\item the main Spoofax meta-languages: SDF2, SDF3, Stratego, NaBL, NabL2, Statix, DynSem, Dynamix;
\item DSLs used in the production of various software systems: currently WebDSL, IceDust, Green-Marl, PGQL, and LeQuest;
\item languages used in Computer Science courses: MiniJava, Jasmin, Tiger, PAPL;
\item demonstration languages provided in MetaBorg \cite{MetaBorg}: SIMPL, QL/QLS, Grace, a subset of Go, MetaC, and Pascal; and
\item specification languages developed for other purposes, such as CBS \cite{Mosses2019SME} (a meta-language for component-based semantics).
\end{itemize}
Spoofax users can exploit the hyperlinks between definitions and references to navigate local clones of code repositories in the above languages; but when accessing the same repositories online, those hyperlinks are not available.

\subsection{Browsing online code repositories}

Suppose that we find a code repository online, and we would like to browse it with precise name-based code navigation using the Spoofax language workbench. Spoofax can only be used to browse local files, so we need to download or clone the repository. We also need to find and download a Spoofax language project for each language used in the code repository (except for the main Spoofax meta-languages). After building those language projects in Spoofax, we can finally browse the cloned repository with precise code navigation. 

The necessity of downloading the code repository and the required language projects surely discourages browsing. Moreover, the Spoofax language workbench is currently available only for use in Eclipse and IntelliJ IDEA; users not familiar with either of those IDEs may be reluctant to install them just for browsing some code repository. And users behind company firewalls might not even be allowed to install such 3rd-party software as Spoofax.

Ideally, an online code repository would support precise name-based navigation in standard web browsers, \emph{without} the need to download any files or install new software. The rest of this paper explains how to do that, using Spoofax itself:
\begin{itemize}
\item Section~\ref{sec:CBS-beta} gives an overview of a repository that supports precise name-based navigation from references to definitions in web browsers.
\item Section~\ref{sec:CBS} presents the Spoofax language project that created the same name-based navigation for local use.
\item Section~\ref{sec:Generation} explains how that navigation is made available in web browsers, using a toolchain that combines Spoofax with some standard applications to create `hyperlinked twins' of unlinked code repositories.
\item Section~\ref{sec:Related} relates the Spoofax-based approach to some other approaches that provide name-based code navigation in web browsers.
\item Section~\ref{sec:Conclusion} concludes with suggestions for further development of the presented approach.
\end{itemize}


\section{The CBS-beta Repository}
\label{sec:CBS-beta}

CBS \cite{CBS-beta} is a meta-language for component-based semantics, developed in the PLanCompS project \cite{PLanCompS}. See previous publications \cite{Churchill2014RCS,Mosses2019SME,Binsbergen2019ECS,Binsbergen2016TSC} for motivation, foundations, and explanations of CBS. The CBS-beta repository provides a beta-release of language specifications using CBS. It has two main parts
\begin{description}
\item[Funcons-beta:] A proposal for an initial library of so-called fundamental programming constructs (funcons).
\item[Languages-beta:] Examples of language specifications based on the funcons in Funcons-beta.
\end{description}
(Two further parts are marked as “unstable”, and still being developed.)

\subsection{CBS specifications} 

Funcons \cite{Mosses2021FCP} are \emph{reusable} components: the same funcon can be used, unchanged, in the specifications of many different languages. Funcons correspond closely to concepts of high-level programming languages such as data and control flow, scopes of bindings, mutable variables, streams, abrupt termination, procedural abstraction, etc.

A funcon definition declares the name of the funcon, and specifies its signature: the types of its arguments (if any), and of the value(s) that it returns. The definition also specifies small-step operational semantics rules for evaluating the funcon.

Funcon names are strongly suggestive of the corresponding concepts. The Funcons-beta library defines about 400 funcons. Many of the funcon names are quite long, but have short aliases  (e.g., “\textsf{allocate-initialised-variable}” has the alias “\textsf{alloc-init}”). 

Funcons are to have \emph{fixed} definitions, so \emph{no version control} will be needed for their safe reuse in CBS language specifications. Crucially, adding new funcons does not require any changes to the definitions of existing funcons, thanks to the use of a modular variant of structural operational semantics (MSOS) \cite{Mosses2004MSO} in CBS. 

Apart from illustrating the use of CBS, the aim of the beta-release of CBS and its initial library of funcons was to allow a public review of the definitions, and subsequent adoption of suggestions for improvement, before their finalisation in a full release. The funcon definitions in Funcons-beta have all been validated by empirical testing. The Funcons-beta library has also been found useful in the iCoLa meta-language for incremental (meta-)programming \cite{Frolich2022ICM}.

A language specification in CBS resembles a conventional denotational semantics: it defines the language syntax by a context-free grammar, and it defines semantic functions -- mapping phrases in the language to their denotations -- by equations. In CBS, the denotations are simply compositions of funcons, instead of higher-order functions.

Languages-beta provides the beta-release of five examples of language specifications in CBS, based on the funcon definitions in Funcons-beta. The examples range from a small introductory example to a substantial sub-language of OCaml.

Each language specification is independent, and (implicitly) imports all the funcons that it references. Regarding name resolution, the Funcons-beta library corresponds to a single module or package, and users do not need to be aware of the internal file structure of the library.

When browsing funcon definitions and language specifications, precise name-based navigation from references to definitions is essential. Section~\ref{sec:CBS} presents a Spoofax language project for CBS that supports such navigation when using Spoofax to browse a local clone of CBS-beta. Table~\ref{tab:scope-spoofax} shows how the definition of the funcon ``\textsf{scope}'' looks in Spoofax. Apart from the first (defining) occurrence of ``\textsf{scope}'', all the names in it are references, equipped with hyperlinks to the respective definitions.

\begin{table}
\caption{Browsing a local clone of the CBS-beta repository in Spoofax.}
\label{tab:scope-spoofax}
\includegraphics[width=\textwidth]{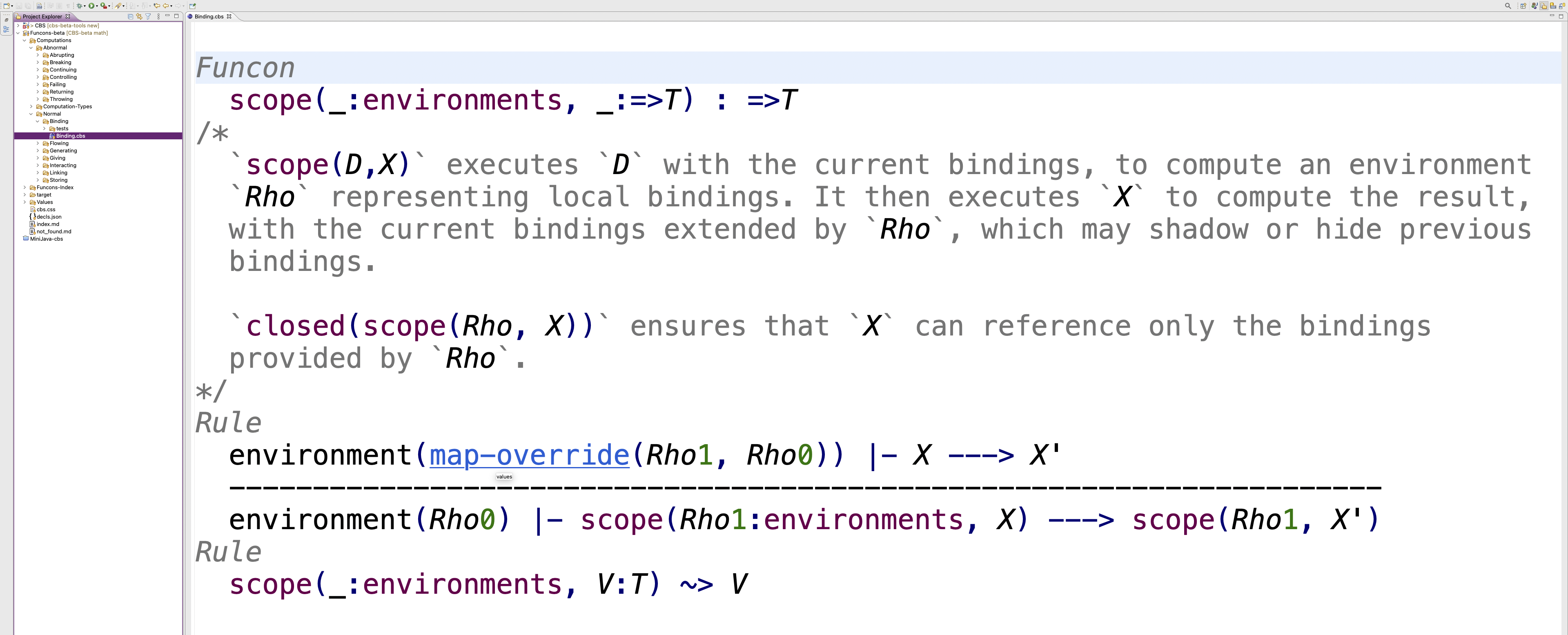}
\end{table}

\subsection{Browsing CBS-beta online}

The CBS sources of the specifications in CBS-beta are available in a repository on GitHub. Table~\ref{tab:scope-github} shows how the same definition as in Table~\ref{tab:scope-spoofax} looks when browsing the CBS sources on GitHub. The text has no hyperlinks, with no support at all for name-based navigation.

However, the CBS-beta repository also contains the sources of an associated website, which includes \emph{hyperlinked twins} of each CBS source file. This website is served by GitHub Pages at \url{https://plancomps.github.io/CBS-beta/}.

The hyperlinked twins of a CBS source file are in the following formats.
\begin{description}
\item[PLAIN:] In this format, the web page displays a verbatim copy\footnote
{The PLAIN format deviates from a verbatim copy by using different font styles.}
of the source file. Table~\ref{tab:scope-plain} shows how the same definition as before looks in the PLAIN format. Names are highlighted: different colours distinguish between names of funcons, syntax sorts, semantic functions, and meta-variables. References are hyperlinked to declarations.
\item[PRETTY:] In this format, the web page displays CBS with mathematical typography (as in published articles about CBS). Table~\ref{tab:scope-pretty} shows how the same definition as before looks in the PRETTY format. As in the PLAIN format, names are highlighted, and references are hyperlinked to declarations.
\item[PDF:] This format is simply a PDF rendering of the PRETTY web page. Table~\ref{tab:scope-pdf} shows how the same definition as before looks in the PDF format.
\end{description}
The PLAIN and PRETTY hyperlinked twins of a CBS source file include links to each other, to the PDF twin, and to the source file on GitHub.

CBS specifications may include informal text as comments, with embedded references to names. In the source files, comments are enclosed in \verb`/*…*/`, and embedded references in back-ticks; in the hyperlinked twins, the comments are displayed as running text, and the formal CBS specifications are displayed as code blocks.

Section~\ref{sec:Generation} explains how Spoofax is used to generate the hyperlinked twins from the CBS source files.

\begin{table}
\caption{Browsing the CBS-beta repository on GitHub.}
\label{tab:scope-github}
\includegraphics[width=\textwidth]{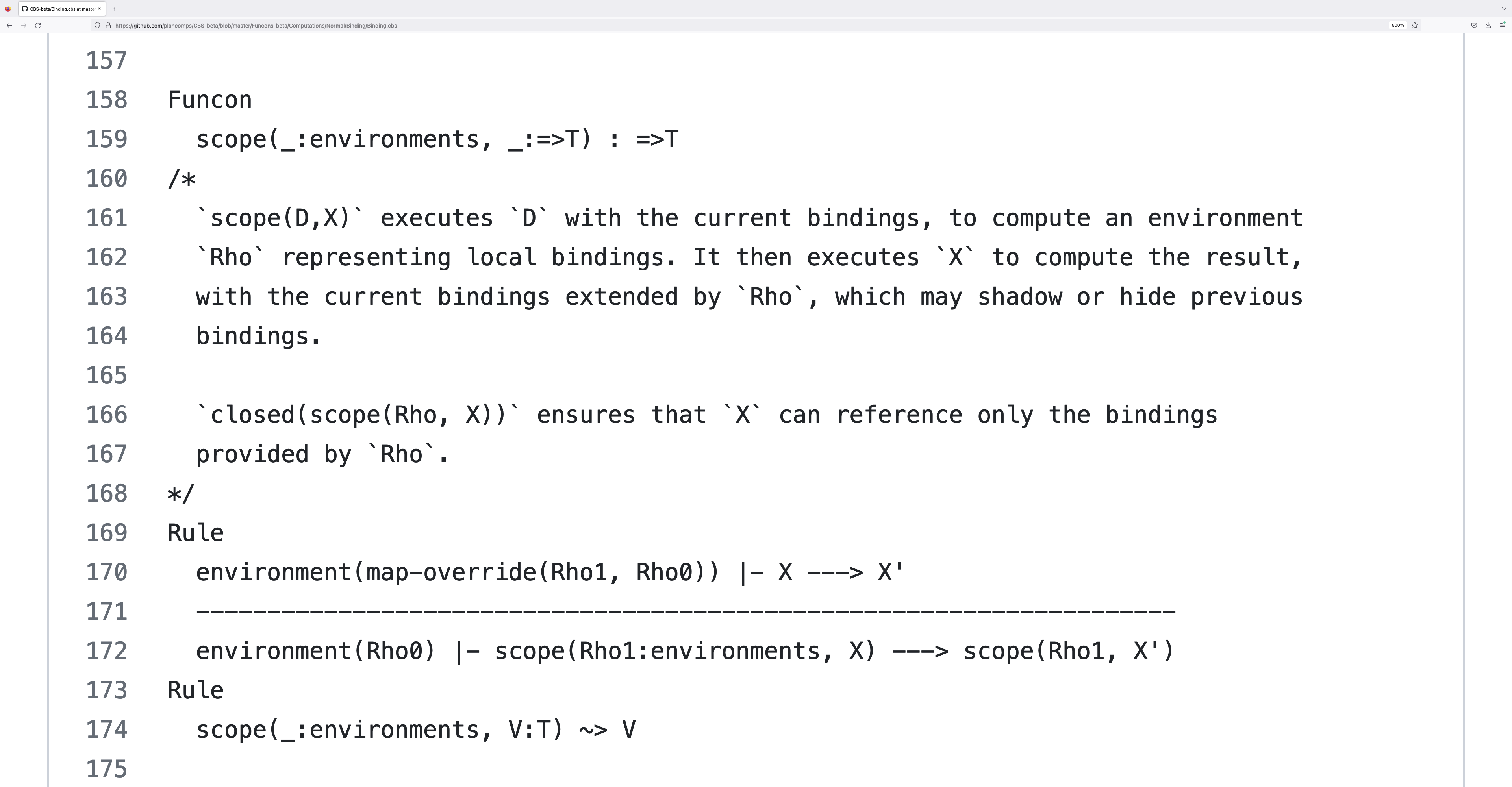}
\end{table}

\begin{table}
\caption{Browsing a PLAIN hyperlinked twin.}
\label{tab:scope-plain}
\includegraphics[width=\textwidth]{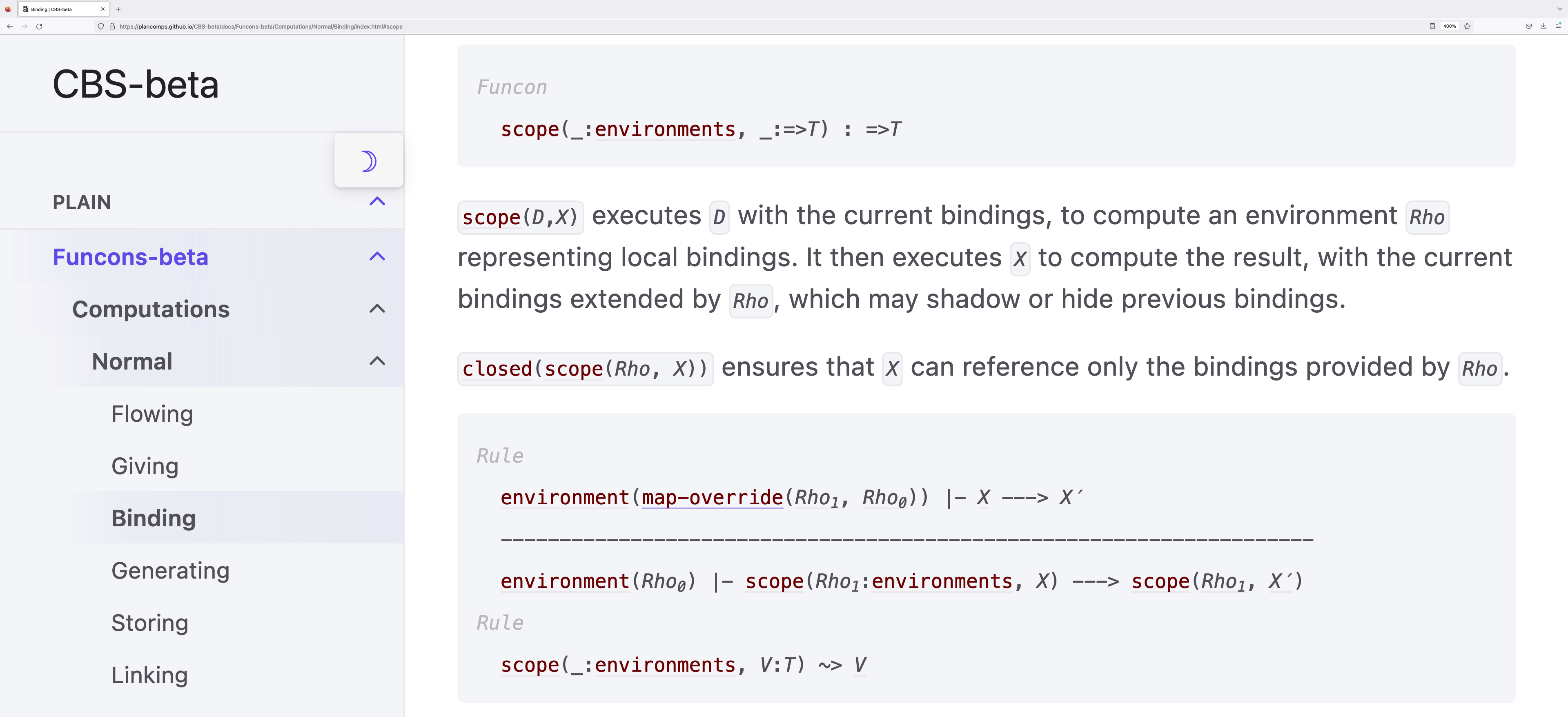}
\end{table}

\begin{table}
\caption{Browsing a PRETTY hyperlinked twin.}
\label{tab:scope-pretty}
\includegraphics[width=\textwidth]{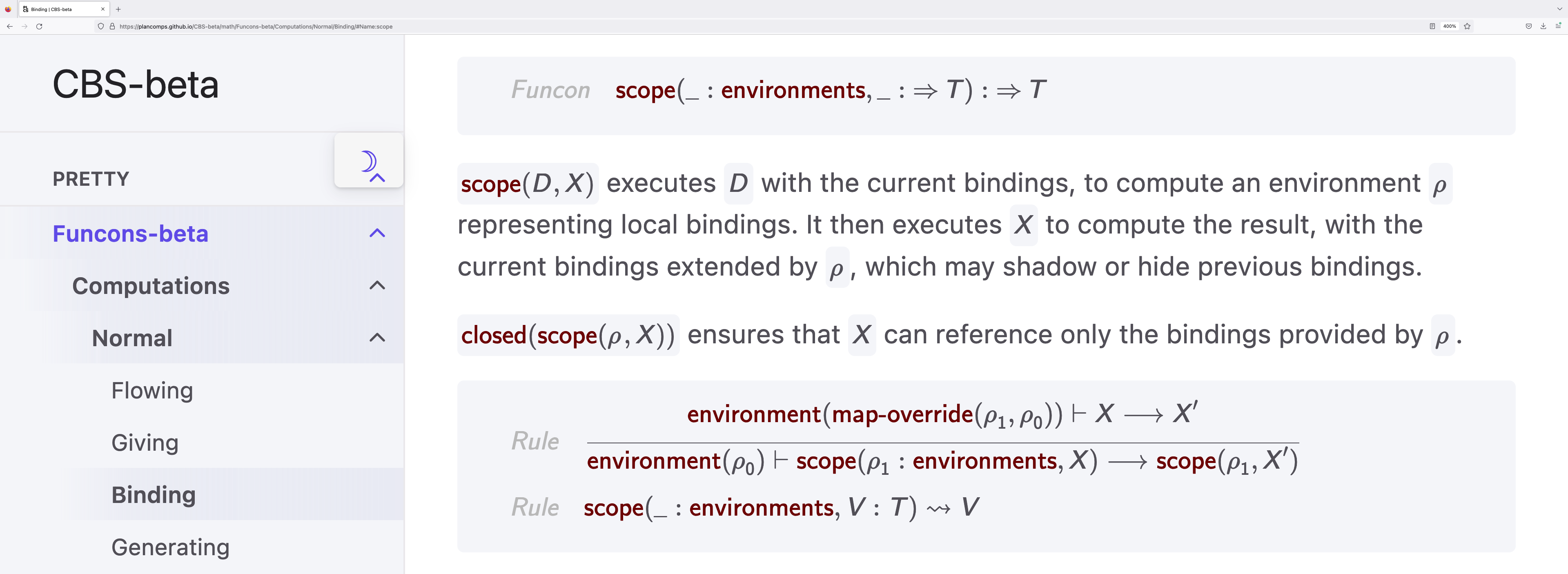}
\end{table}

\begin{table}
\caption{Browsing a PDF hyperlinked twin.}
\label{tab:scope-pdf}
\includegraphics[width=\textwidth]{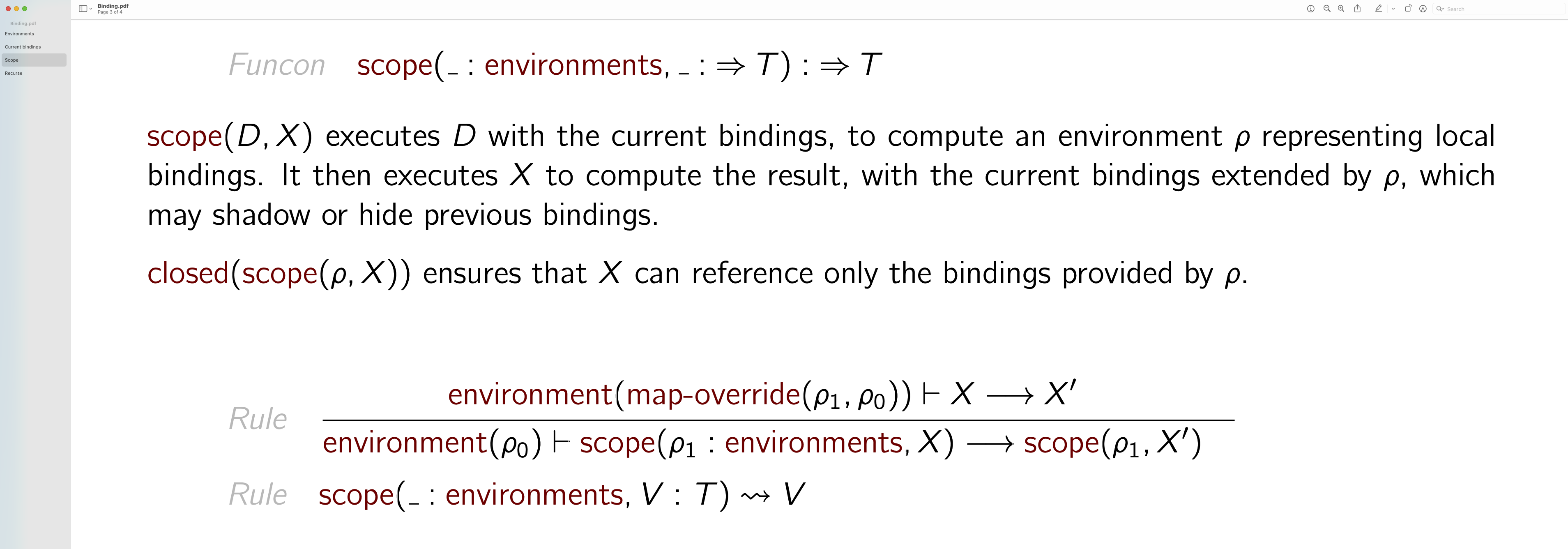}
\end{table}

\section{A Spoofax Language Project for CBS}
\label{sec:CBS}

The author has developed a Spoofax language project for CBS. It specifies the syntax of CBS using the meta-language SDF3. Spoofax generates a parser from the SDF3 specification. When a CBS source file is opened, Spoofax creates an editor window for the text, and automatically re-parses the text each time it is edited.

Spoofax uses syntax highlighting to display well-formed phrases; when an edit introduces an error, Spoofax moves the cursor to the source of the error. Spoofax also warns about any phrases that have ambiguous parses.

The syntax of CBS is simpler than that of typical high-level programming languages, and its specification in SDF3 was reasonably straightforward. Unusually, comments are included in ASTs, and formal terms can be embedded in comments (enclosed in back-ticks, as in Markdown).

The CBS language project specifies name resolution using the meta-language NaBL2 \cite{Neron2015TNR,Antwerpen2016CLS}. Name resolution in CBS involves checking that all referenced names of each sort (syntax, semantics, funcons) are declared uniquely. Type analysis checks that semantic functions are applied only to the sort of syntax argument specified in their signatures, and that funcons are applied only to the expected number of arguments. However, the analysis of a CBS specification involves the analysis of the entire funcons library, and the implementation of NaBL2 is non-incremental, so re-analysis needs to be suspended while editing even a small CBS source file. (See Section~\ref{sec:Conclusion} for how to address this drawback.)

The CBS language project also supports generation of parsers and translators from language specifications. It transforms the context-free grammar of a language specification in CBS to the corresponding SDF3 grammar; it transforms the equations defining the semantic functions to corresponding rewrite rules in Stratego. The resulting language project can parse programs in the specified language, then translate the programs to funcon terms.

\begin{figure}
\includegraphics[width=\textwidth]{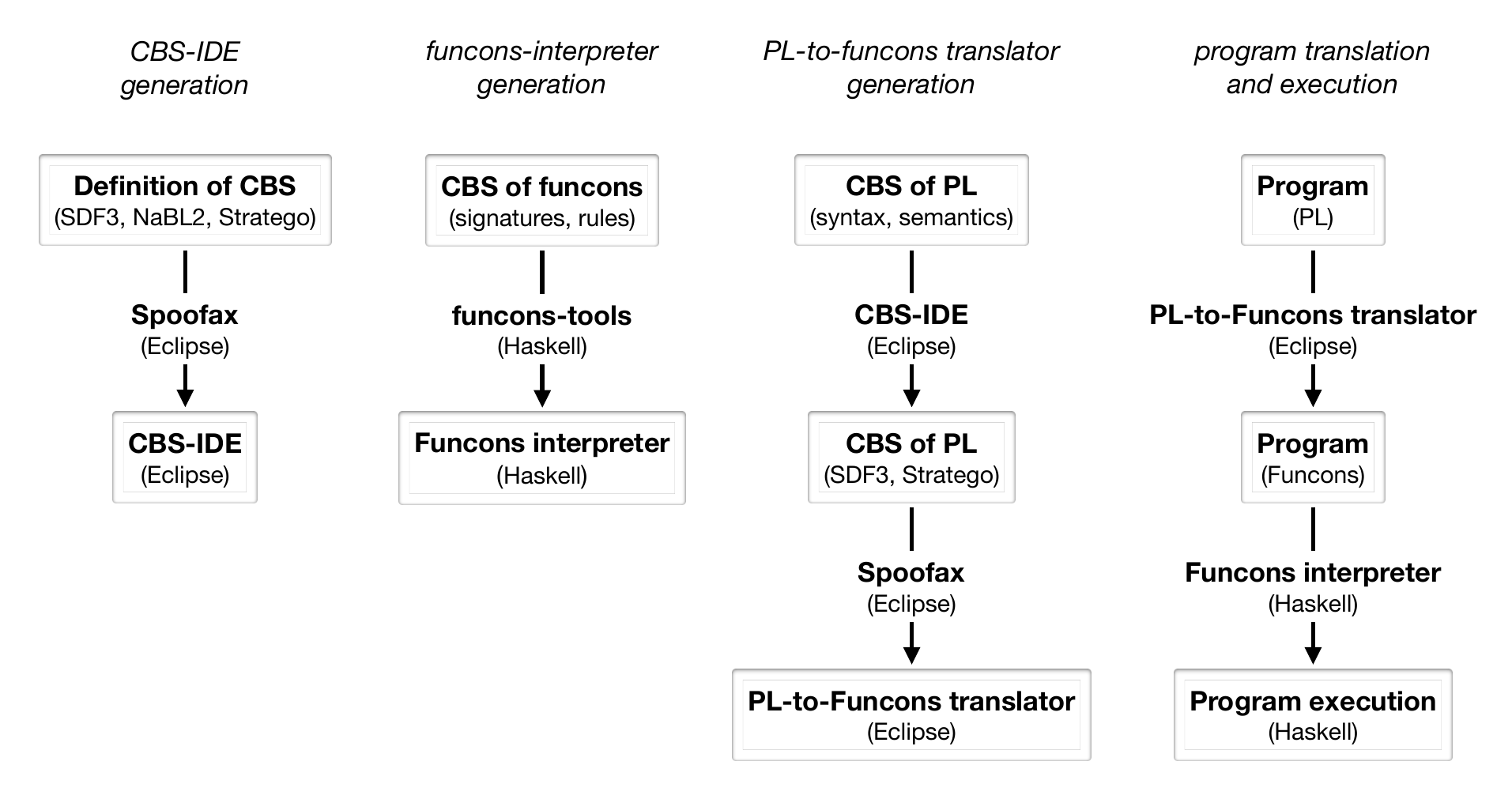}
\caption{Generation and use of a CBS-IDE based on Spoofax and external tools \cite{Mosses2019CFL}.}
\label{fig:tools}
\end{figure}

External tools (implemented originally in Prolog, and subsequently in Haskell \cite{Binsbergen2019ECS}) support evaluation of funcon terms according to the rules that define the funcons, thereby testing whether the rules specify the expected results. Moreover, the combination of these external tools with the Spoofax language project generated from a language specification allows programs in the specified language to be run according to their translation to funcons; this tests the translation rules as well as the rules from the funcon definitions. Figure~\ref{fig:tools} gives an overview of the tool chain; see reference \cite{Mosses2019CFL} for further details. \emph{Running the semantics} on suites of test programs, using artefacts generated automatically from the semantics, can reveal subtle errors, as well as eliminating trivial mistakes \cite{Klein2012RYR}.

\section{Generation of Hyperlinked Twins}
\label{sec:Generation}

This section explains how to provide the same name-based code navigation in web browsers as in Spoofax. It uses a toolchain that combines Spoofax with some standard applications to produce \emph{hyperlinked twins} of CBS source files: web pages displaying the same content as the source files, but with the addition of hyperlinks from references to declarations. The web pages also add syntax highlighting, corresponding to that provided by Spoofax when browsing CBS source files locally.

The CBS language project generates hyperlinked twins of CBS source files in three formats: PLAIN, PRETTY, and PDF. For each format, Stratego rules specify a recursive traversal of the analysed AST of the source file, generating strings in the \verb`kramdown` markup language (a variant of Markdown) \cite{kramdown}. The NaBL2 API for Stratego provides strategies to test whether a node of the AST is a declaration or a reference; and when name resolution has determined the declaration to which a reference refers, the API also provides the path of the file that contains the declaration.

Recall that CBS specifications may include informal text as comments. In the source files, comments are enclosed in \verb`/*…*/`; the hyperlinked twins display comments as running text, and \verb`kramdown` determines the layout. Embedded references to names in comments become (highlighted) hyperlinks to declarations. The formal parts of the CBS specifications are displayed as code blocks.

\subsection{PLAIN format}

The PLAIN format preserves the layout (indentation, line breaks) of the CBS specification by enclosing it in a pre-formatted HTML element (\verb`<pre>`), which is rendered by browsers with a fixed-width font. References and declarations in CBS generate anchor elements (\verb`<a>`) in HTML\@. Syntax highlighting is produced by HTML span elements. The static site generator Jekyll \cite{Jekyll} renders the generated \verb`kramdown` file (prefixed with some meta-data) on the CBS-beta website.

The PLAIN format shows how to write CBS in source files, and the correspondence between a source file and the hyperlinked web page is direct. However, CBS is based on mathematical notation from denotational and operational semantics, and publications about CBS generally display specifications with mathematical typography. The PLAIN format represents an inference rule by a sequence of dashes between the the premises and the conclusion; and it represents a labelled transition by enclosing the label in \verb`--` and \verb`–>`. Some of the other approximations of mathematical symbols by ASCII characters are similarly indirect, as well as inelegant. The PRETTY and PDF formats address those issues.

\subsection{PRETTY format}

For the PRETTY format, the Spoofax language project for CBS generates {\LaTeX} math-mode markup from the analysed ASTs of CBS source files. The \verb`kramdown` variant of Markdown allows such blocks, but leaves their rendering to math typesetting libraries such as KaTeX~\cite{KaTeX} and MathJax \cite{MathJax}.

To avoid dependence on low-level details of the {\LaTeX} commands supported by KaTeX and MathJax, the author has developed CBS-LaTeX \cite{CBS-LaTeX}, a small {\LaTeX} package for CBS specifications. When CBS is marked up using the commands defined by the package, {\LaTeX} formatting produces mathematical typography, suitable for inclusion in published articles. CBS-LaTeX also includes KaTeX and MathJax configurations that produce similar-looking results from the same {\LaTeX} mark-up when embedded in web pages. 

Markup using CBS-LaTeX is quite low-level; this makes it easy to adjust the layout to fit the intended page width, but tedious to write. The markup generated by the CBS language project includes line breaks in places where they should enhance readability; it also respects the line breaks in formulae in the source files.

\subsection{PDF format}

For the PDF format, \verb`kramdown` takes the Markdown with {\LaTeX} math blocks used for the PRETTY format, and converts the Markdown to text-mode {\LaTeX}. Using the CBS-LaTeX package, \verb`pdflatex` then produces a PDF document where the layout and formatting match the rendering of the PRETTY web page in the browser.

\subsection{Offline generation}

Although the definitions of funcons are to be fixed (after the finalisation of Funcons-beta) their grouping in files may change. Moreover, the informal comments that motivate and explain the funcons are not definitive, and can change. The Spoofax language project for CBS provides buttons to update the generated Markdown files when needed, and GitHub Pages automatically rebuilds the CBS-beta website when changes are pushed to the CBS-beta repository. However, Spoofax can be used offline: the application Sunshine2 \cite{spoofax-sunshine} can open Spoofax in Eclipse, build a specific project, and execute the actions attached to buttons. Offline regeneration of all outdated files has been automated using a Makefile.

\begin{figure}
\centering
\includegraphics[width=0.9\textwidth]{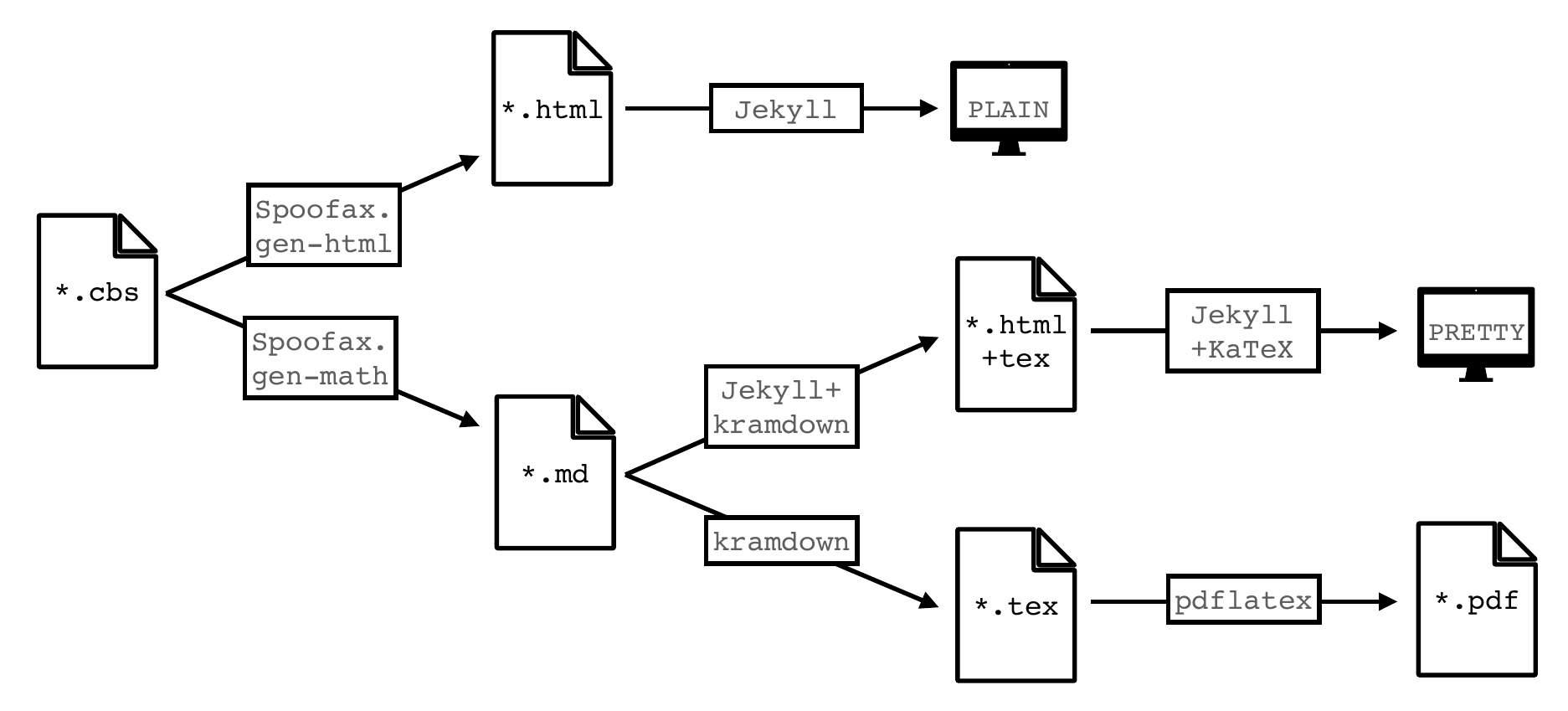}
\caption{Generation of hyperlinked twins using Spoofax and external tools.}
\label{fig:twins}
\end{figure}

Figure~\ref{fig:twins} summarises the toolchains used to generate the PLAIN, PRETTY, and PDF hyperlinked twins of CBS source files.

\section{Related Approaches}
\label{sec:Related}

The generation of websites from \href{https://agda.readthedocs.io/en/latest/tools/literate-programming.html}{Literate Agda} specifications provided the initial inspiration for a hyperlinked twin of the CBS source files. In particular, the online book \emph{Programming Language Foundations in Agda} \cite{Wadler2022PLF} was generated from Literate Agda sources, with name references hyperlinked to definitions. Agda is an indentation-sensitive language, so the generated HTML respects layout in the same way as the PLAIN format on the CBS-beta website. Literate Agda allows informal text to be marked up in Markdown. One difference is that Agda source files make extensive use of Unicode characters, in contrast to the ASCII approximation to mathematical notation used in CBS source files. The implementation of Literate Agda uses \verb`pandoc` to convert Markdown to HTML, analogously to how the CBS-beta website uses \verb`kramdown`.

GitHub currently supports \emph{search-based} code navigation for files in 10 languages: C\#, CodeQL, Elixir, Go, Java, JavaScript, PHP, Python, Ruby, and TypeScript. Searching for the definition of a referenced name may return irrelevant results; similarly when searching for all references to a particular definition of a name. 

GitHub also supports \emph{precise} code navigation for Python. The name resolution is based on stack-graphs \cite{Creager2021ISG}, which are closely related to the scope graphs used in Spoofax. Understandably, GitHub is focusing its support for precise code navigation on languages that are widely used in its repositories. The cited reference states:
\begin{quote}
Over the coming months, we will add stack graph support for additional languages, allowing us to show precise code navigation results for them as well. Our stack-graphs library is open source and builds on the \href{https://tree-sitter.github.io/}{\emph{Tree-sitter}} ecosystem of parsers. We will also be publishing information on how language communities can self-serve stack graph support for their languages, should they wish to.
\end{quote}
In principle, it should be possible to migrate the grammar for CBS from SDF3 to \emph{Tree-sitter}, and the name resolution rules for CBS from NaBL2 to stack-graphs. But it appears that doing that would not provide code navigation in CBS repositories on GitHub.

An alternative approach would be to implement CBS support for the Language Server Protocol (which is not currently incorporated in Spoofax). Then web-enabled IDEs (e.g., Visual Studio Code \cite{VSCode}) would be able browse CBS repositories both locally and online. Name resolution for CBS would need to be implemented in a code indexing format such as LSIF or SCIP \cite{SCIP}. That might be straightforward for CBS (due to the uniqueness and global visibility of funcons in Funcons-beta) but perhaps not for the Spoofax meta-languages and the other languages that have already been specified in Spoofax.


\section{Conclusion and Future Work}
\label{sec:Conclusion}

Spoofax has been used in toolchains to generate the PLAIN, PRETTY, and PDF hyperlinked twins of the specifications in the CBS-beta repository. It could be used in the same way for any other repository of specifications in the CBS-beta language, after setting up a Jekyll website in it.

In the CBS-beta repository, each language in Languages-beta is a separate Spoofax language project, with a symbolic link to the Funcons-beta project. On GitHub, other repositories could include Funcons-beta as a submodule, to make the funcon definitions locally available and avoid the need for inter-repository name resolution.

It should be straightforward to generate PLAIN hyperlinked twins of specifications written in the Spoofax meta-languages (SDF3, Statix, etc.), since name resolution for those languages has already been specified in Spoofax. The tree-to-string traversal is specified generically in Stratego for nodes that define or reference names; adding syntax highlighting involves inserting further HTML markup for the relevant nodes, but other nodes generate strings uniformly. To generate PRETTY and PDF twins would require transformation to combinations of Markdown and {\LaTeX} math blocks.

The techniques presented here might be useful for online code navigation also in small programming languages (including DSLs) and specification languages when their syntax and name resolution can be easily specified in Spoofax. For other languages, the alternative techniques to support code navigation discussed in Section~\ref{sec:Related} may be preferred. 

To be able to jump straight to the definition of a funcon from all references to its name is the most important feature of name-based navigation in CBS-beta specifications. Currently, the generated hyperlinked twins do not support finding all references to definitions. The list of all references to a definition is available in the analysed AST, and displayed by Spoofax when browsing the definition; it could be displayed in the same way in the PLAIN and PRETTY twins, but it is unclear how to display it in the PDF twin.

The author is planning to re-specify name resolution for CBS-beta in the Statix meta-language. Statix has an incremental implementation, which will avoid repeatedly re-analysing the Funcons-beta project while editing a language specification. This should remove the main source of inefficiency with interactive use of the current Spoofax language project for CBS-beta, in preparation for a release of the CBS-beta language project as an Eclipse plugin.

\bibliography{EVCS}

\end{document}